# SPECTRA FOR THE PRODUCT OF GAUSSIAN NOISES

**Laszlo Bela Kish[1), Robert Mingesz[2)], Zoltan Gingl[2)], Claes-Goran Granqvist[3)]**

1) Texas A&M University, Department of Electrical and Computer Engineering, College Station, TX 77843-3128, USA
2) Department of Technical Informatics, University of Szeged, Árpád tér 2, Szeged, H-6701, Hungary
3) Department of Engineering Sciences, The Ångström Laboratory, Uppsala University, P.O. Box 534, SE-75121 Uppsala, Sweden

**Abstract**

Products of Gaussian noises often emerge as the result of non-linear detection techniques or as a parasitic effect, and their proper handling is important in many practical applications, including in fluctuation-enhanced sensing, indoor air or environmental quality monitoring, etc. We use Rice's random phase oscillator formalism to calculate the power density spectra variance for the product of two Gaussian band-limited white noises with zero-mean and the same bandwidth $W$. The ensuing noise spectrum is found to decrease linearly from zero frequency to $2W$, and it is zero for frequencies greater than $2W$. Analogous calculations performed for the square of a single Gaussian noise confirm earlier results. The spectrum at non-zero frequencies, and the variance of the square of a noise, is amplified by a factor two as a consequence of correlation effects between frequency products. Our analytic results is corroborated by computer simulations.

## 1. Introduction

Noise entering non-linear systems gives rise to higher-order product terms, and the smallest non-linear order is the second order. Such second-order components can be the product of two independent noises in the system, or they can be the square of a single noise. These non-linear noise components often carry important information about the system. For example, fluctuation-enhanced sensing (FES) [1-8] utilizes the information provided by the sensor noise, and this information can be significantly richer [1-3] than the sensory information of the deterministic signal component used for classical sensing.

In FES, the most frequently used method for extracting sensory information employs the analysis and pattern recognition of the power spectral density (PSD or "noise spectrum"), but higher-order techniques and other special methods—such as zero-crossing analysis and correlation studies—have also been successfully tested [2,6,7].

FES has a demonstrated capability to detect and identify harmful gases [4] and odors of dangerous bacteria [5] even at low concentrations. This fact, and the high information content in the FES technique, makes it possible to distinguish between many different chemical compositions, which is highly relevant for environmental and indoor air quality monitoring since there is a huge variety of potentially harmful gases and vapors [8,9].

Knowledge and analysis of the specific noise spectra that can occur in a FES device are important for its optimal design and function, and the present paper deduces noise spectra for a common and practical case: that of the product of two independent Gaussian band-limited white noises with the same bandwidth $W$. Analogous information is obtained also for the square of a single Gaussian noise of the same character and the results are in agreement with know results about this particular situation. We use Rice's rigorous method [10] based on random phase oscillators and modulation components, which is valid for Gaussian noises. The triangular spectral shapes found by us are in accordance with Bennett's heuristic non-exact method for cross-modulation products of different speech channels [11].

**2. Rice's random-phase oscillator model**

Consider two Gaussian noises of zero mean with amplitudes $U_1(t)$ and $U_2(t)$ and power spectral densities $S_1(f)$ and $S_2(f)$, respectively. Rice's random-phase oscillator representation of these noises can be used to synthesize the amplitude of a Gaussian noise with power spectral density $S_0(f)$ as the sums of oscillators. The frequency scale is divided into infinitesimally small intervals with bandwidth $f_0$, and the noise in each interval is represented by an oscillator with oscillation frequency $kf_0$,

amplitude $a(f) = a(kf_0) = \sqrt{2 f_0 S_0(f)} = \sqrt{2 k f_0 S_0(kf_0)}$, and

random phase $\varphi_k$ uniformly distributed over the $[0, 2\pi)$ interval.

Thus the two noises can be synthesized as

$$U_1(t) = \lim_{N \to \infty} \sum_{j=1}^{N} a_j \sin(2\pi j f_0 t + \varphi_j) \text{ and } U_2(t) = \lim_{N \to \infty} \sum_{k=1}^{N} b_k \sin(2\pi k f_0 t + \varphi_k), \tag{1}$$

where $a_j = \sqrt{2 f_0 S_1(jf_0)}$ and $b_k = \sqrt{2 f_0 S_2(kf_0)}$.

Then the amplitude of the product $U(t)$ of the two noises is given as

$$U(t) = \gamma U_1(t) U_2(t) = \gamma \lim_{N \to \infty} \sum_{j=1}^{N} a_j \sum_{k=1}^{N} b_k \sin(j 2\pi f_0 t + \varphi_j) \sin(k 2\pi f_0 t + \varphi_k) =$$

$$= \gamma \lim_{N \to \infty} 0.5 \sum_{j=1}^{N} a_j \sum_{k=1}^{N} b_k \cos[(j+k) 2\pi f_0 t + \varphi_j + \varphi_k] + \tag{2}$$

$$+ \gamma \lim_{N \to \infty} 0.5 \sum_{j=1}^{N} a_j \sum_{k=1}^{N} b_k \cos[(j-k) 2\pi f_0 t + \varphi_j - \varphi_k]$$

where $\gamma = 1/\text{Volt}$ is the transfer coefficient of a hypothetical multiplier device providing a Volt unit also for the product. The first sum represents the combination-frequencies obtained as the sum of frequencies ("additive" frequency products) while the second sum represents the difference-frequencies ("subtractive" frequency products).

Three facts must be kept in mind when determining the PSD:

(*i*) the power of components with different frequencies are additive,

(*ii*) the power of components with the same frequency but random phase are also additive after the ensemble averaging of an infinite number of other components in the immediate vicinity of that frequency, and

(*iii*) the power of frequency components with the same frequency and same phase add up in a synergetic way because of their total correlation; the amplitudes are additive and the power scales with the square of the amplitudes

**3. Spectrum for independent noises**

In the case of independent $U_1(t)$ and $U_2(t)$, both (*i*) and (*ii*) are satisfied, and thus the result of Eq. (1) is free of correlation effects. Fig. 1 shows the shape of the ensuing noise spectrum: The spectral bands belonging to a given *j* parameter (*cf.* the horizontal rectangles) are summed to determine the shape of the related noise spectrum. Left-upper part of Fig. 1 shows the "additive" frequency products where *j* is increasing from bottom to top while *k* runs horizontally, and left-lower part shows the "subtractive" frequency products where *j* is increasing from top to bottom. Right-upper part of Fig. 1 shows the resulting spectra, which are proportional to the frequency-dependent vertical thickness of the "additive" and "subtractive" spectral patterns in the left-hand part of the figure. It is seen that this procedure yields triangular spectra with center frequencies at *W* and 0, respectively. Right-lower part of Fig. 1 shows the final result, which is the sum of the two triangular spectra (negative frequencies are flipped to positive) which gives rise to a linearly decreasing spectrum going from $S(0)$ at zero frequency to zero at $2W$ frequency, *i.e.*,

$$S(f) = S(0)\left(1 - \frac{f}{W}\right) \text{ for } 0 \le f \le W; \text{ elsewhere } S(f) = 0 \ . \tag{3}$$

The magnitude of $S(0)$ will be determined in the next section.

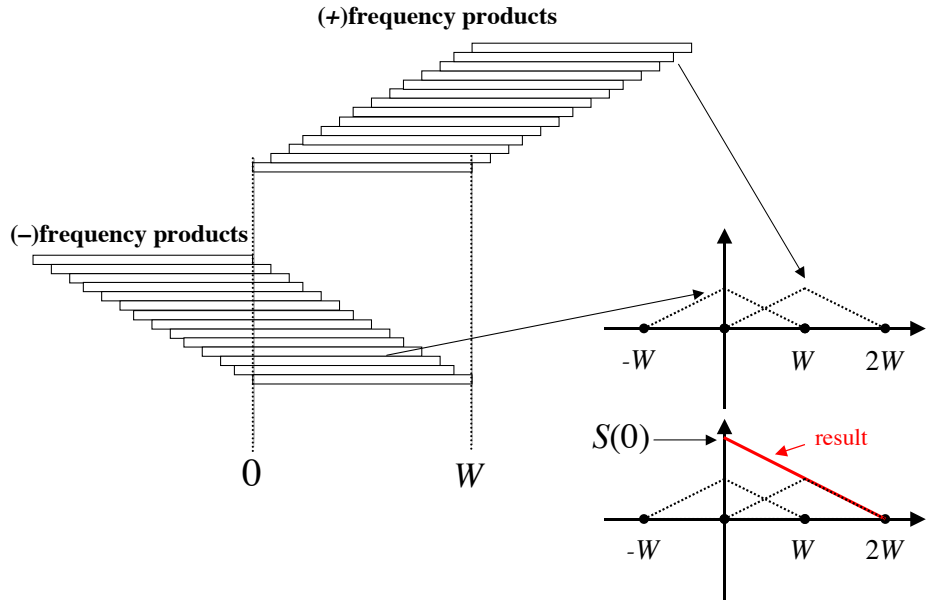

Fig. 1. Determination of the shape of the noise spectrum by summing spectral bands (see the text for explanation).

### 4. Spectrum and variance for independent noises

The variance of the product of two independent random variables with random mean value is the product of the original variances so that

$$\langle U^2(t)\rangle = \gamma^2 \langle U_1^2(t)\rangle \langle U_2^2(t)\rangle \ . \tag{4}$$

The variance is the area below the noise spectrum, and thus

$$\langle U_1^2(t)\rangle = WS_1(f), \ \langle U_2^2(t)\rangle = WS_2(f), \text{ and } \langle U^2(t)\rangle = WS(0) \ . \tag{5}$$

Eqs. (3) to (5) yield the spectrum of the product of two independent noises according to

$$S(f) = \gamma^2 W S_1 S_2 \left(1 - \frac{f}{W}\right) \text{ for } 0 \leq f \leq W; \text{ elsewhere } S(f) = 0 . \tag{6}$$

### 5. Spectrum and variance for the square of a noise

A special and well-studied situation occurs when a noise is multiplied by itself. For the square of a noise $U_1(t)$, conditions (*i*) and (*ii*) are not always satisfied because for each non-zero frequency combination in Eq. (2) there will be a sum of two identical products instead of their random phase sum. This correlation leads to a doubling of the spectrum obtained in Eq. (6). Furthermore the spectrum is infinite at zero frequency as a consequence of the non-zero DC component emerging from the relevant correlations. Thus one obtains for non-zero-frequencies that

$$S(f) = 2\gamma^2 W S_1 S_2 \left(1 - \frac{f}{W}\right) \text{ for } 0 < f \leq W; \quad S(f) = 0 \text{ for } f < 0 \text{ and } f > W . \tag{7}$$

It then follows that the variance of the noise component (the AC component) of the square of the noise $U_1(t)$ is twice the result given in Eq. (4), *i.e.*,

$$\left\langle U_{AC}^2(t) \right\rangle = \gamma^2 \left\langle \left[U_1^2(t) - \left\langle U_1^2(t) \right\rangle\right]^2 \right\rangle = 2\gamma^2 \left\langle U_1^2(t) \right\rangle^2 . \tag{8}$$

Our results agree with those in a related study [10].

### 6. Computer simulations

Computer simulations were carried out to illustrate the results reported above.

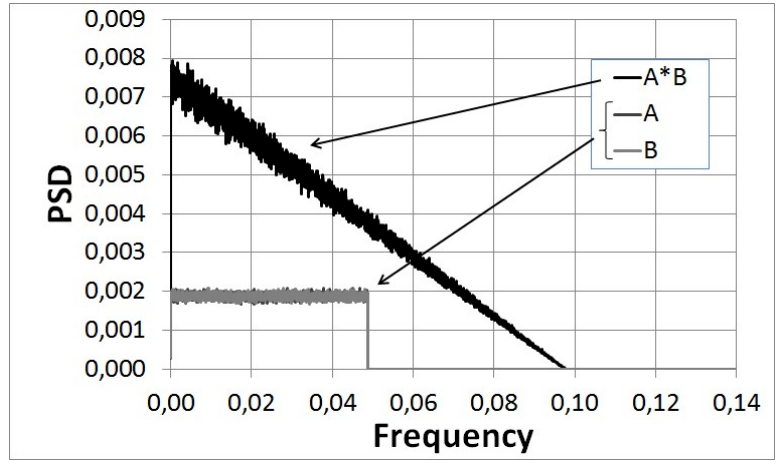

Fig. 2. Computer simulation of power spectral density (PSD). The overlapping input noises (A and B) are band-limited white noises, while the resulting spectrum is the expected linearly decreasing function reaching zero at twice the original bandwidth.

Two independent Gaussian noises were generated with the same spectra and bandwidth, and the PSD and variance of the product of these noises were determined. Fig. 2 shows that the shape of the resulting noise spectrum is in agreement with the shape predicted by Eq. (6). Analogously, the variances of the product of the noises and the AC component of the square of a noise satisfy Eqs. (4) and (8), respectively.

Fig. 3 shows results of a computer simulation of the resulting amplitude distribution function without filtering (sharp peak) and after cutting off the bandwidth at the original bandwidth (Gaussian-like distribution). This simulation was carried out in order to illustrate how the practical situation, with the same bandwidth of the output as that of the input noises, strongly diminishes the non-Gaussian effects. The original, heavily non-Gaussian distribution (Normal Product Distribution [12]) clearly significantly changed toward a much more Gaussian-looking bell-shaped curve.

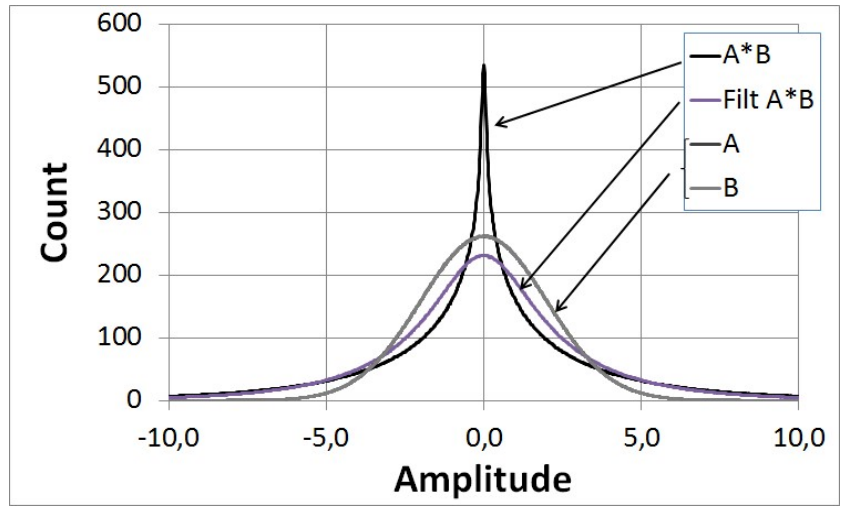

Fig. 3. Amplitude distribution of the original noises (A and B), of the product noise with $2W$ bandwidth (A*B), and of the product noise after filtering to have the same bandwidth $W$ as the original noises (Filt A*B).

## 7. Conclusions

This paper deduced simple results for noise spectra relevant to a practical case: the noise spectrum of the product of two independent Gaussian band-limited white noises with the same bandwidth. We used Rice's method [10] based on random phase oscillators and modulation components, which is valid for Gaussian noises. The triangular spectral shapes found by us are in accordance with Bennett's heuristic non-exact method for cross-modulation products of different speech channels [11]. Analogous results were obtained for the square of a single Gaussian noise of the same character as above. The latter result was known from prior work [10] and largely served as a confirmation of our analysis.

Our results have potential applications in the analysis and design of practical devices for correlators, secure classical communicators, fluctuation-enhanced sensing of indoor air and environmental quality, etc.

**Acknowledgements**

A discussion with Professor L. Stacho is appreciated. Financial support for LBK's visit of Angstrom Laboratory was received from the European Research Council under the European Community's Seventh Framework Program (FP7/2007-2013)/ERC Grant Agreement 267234 ("GRINDOOR"). Work in Szeged was supported by the grant TAMOP-4.2.1/B-09/1/KONV-2010-0005.

# References

[1] Kish, L.B., Vajtai, R., Granqvist, C.G. (2000). Extracting information from noise spectra of chemical sensors: Single sensor electronic noses and tongues. *Sensors and Actuators B* 71, 55–59.

[2] Smulko, J., Ederth, J., Kish, L.B., Heszler, P., Granqvist, C.G. (2004). Higher-order spectra in nanoparticle gas sensors. *Fluctuation and Noise Letters* 4, L597–L603.

[3] Ederth, J., Smulko, J.M., Kish, L.B., Heszler P., Granqvist, C.G. (2006). Comparison of classical and fluctuation-enhanced gas sensing with $Pd_xWO_3$ nanoparticle films. *Sensors and Actuators B* 113, 310–315.

[4] Kish, L.B., Li, Y., Solis, J.L., Marlow, W.H., Vajtai, R., Granqvist, C.G., Lantto, V., Smulko, J.M., Schmera, G. (2005). Detecting harmful gases using fluctuation-enhanced sensing. *IEEE Sensors Journal* 5, 671–676.

[5] Kish, L.B., Chang, H.C., King, M.D., Kwan, C. Jensen, J.O., Schmera, G., Smulko, J., Gingl, Z., Granqvist C.G. (2011). Fluctuation-enhanced sensing for biological agent detection and identification. *IEEE Nanotechnology* 10, 1238–1242.

[6] Gingl, Z., Kish, L.B., Ayhan, B., Kwan, C., Granqvist, C.G. (2010). Fluctuation-enhanced sensing with zero-crossing analysis for high-speed and low-power applications. *IEEE Sensor Journal* 10, 492–497.

[7] Schmera, G., Gingl, Z., Kish, L.B., Ayhan, B., Kwan, Granqvist, C.G., (2010). Separating chemical signals of adsorption-desorption and diffusive processes. *IEEE Sensors Journal* 10, 461–464.

[8] Granqvist, C.G., Azens, A., Heszler, P., Kish, L.B., Österlund L. (2007). Nanomaterials for benign indoor environments: Electrochromics for "smart windows", sensors for air quality, and photo-catalysts for air cleaning. *Solar Energy Materials and Solar Cells* 91, 355–365.

[9] Smith, G.B., Granqvist, C.G. (2010). *Green Nanotechnology: Solutions for Sustainability and Energy in the Built Environment*, CRC Press, Boca Raton, FL, USA.

[10] Rice, S.O. (1944). Mathematical analysis of random noise. *Bell System Technical Journal* 23, 282–332.

[11] Bennett, W.R. (1940). Cross-modulation requirements of multichannel amplifiers below overload. *Bell System technical Journal* 19, 587–610.

[12] Normal Product Distribution, Wolfram Mathworld.
http://mathworld.wolfram.com/NormalProductDistribution.html